\documentstyle[12pt,epsfig]{article}
\newcommand{\gtrsim}{\raisebox{-0.8mm}%
{\hspace{1mm}$\stackrel{>}{{\scriptstyle \sim}}$\hspace{1mm}}}
\newcommand{\lessim}{\raisebox{-0.8mm}%
{\hspace{1mm}$\stackrel{<}{{\scriptstyle \sim}}$\hspace{1mm}}}

\def\thebibliographyss#1{\subsubsection*{References}\list
  {[\arabic{enumi}]}{\settowidth\labelwidth{[#1]}\leftmargin\labelwidth
    \advance\leftmargin\labelsep
    \usecounter{enumi}}
    \def\newblock{\hskip .11em plus .33em minus -.07em}
    \sloppy
    \sfcode`\.=1000\relax}

\begin{document}
\begin{flushright}
LU TP 96--5 \\
MC--TH--96/04 \\
January 1996 
\end{flushright}
\begin{center}
{\bf Parton distributions of real and virtual photons }\footnote{To 
 appear in the proceedings of the Workshop on HERA Physics, Durham,
 England, Sept.\ 1995.}
\bigskip

{\bf T. Sj\"ostrand$^{\, a}$, J.K. Storrow$^{\, b}$ and A. Vogt$^{\, c,}
$}\footnote{On leave of absence from Sektion Physik, Universit\"at 
M\"unchen, D-80333 Munich, Germany}

\medskip
$^a${\sl Department of Theoretical Physics, University of Lund,
Lund, Sweden} 

$^b${\sl Department of Theoretical Physics, University of Manchester, 
Manchester, UK}

$^c${\sl Deutsches Elektronen-Synchrotron DESY, Hamburg, Germany}

\end{center}

{\bf Abstract}

Recent progress on the parton distribution functions of the photon, 
both real and virtual, is briefly reviewed and experimental 
possibilities at HERA are discussed.

{\bf 1. Introduction}

Before the advent of HERA, the almost only experimental information on 
the parton structure of the photon was obtained from studies of the 
structure function of the photon, $F_2^\gamma (x,Q^2)$, in two photon 
collisions at e$^+$e$^-$ colliders. Theoretically, this is a very 
interesting area as, at large $x$ and asymptotically large $Q^2$, the 
parton distribution functions (pdfs) of the photon, and hence 
$F_2^\gamma $, are predicted from perturbative QCD (pQCD) \cite
{Witten,Bard}. However, in the range of $Q^2$ experimentally accessible
at present and in the foreseeable future, some non-perturbative input 
is required. Here different groups make different assumptions, all 
include parameters, and nearly all existing pdfs are constrained by fits
to $F_2^\gamma $ data. There are many competing sets; in sect.\ 2 we 
review the various possibilities, discuss the basic underlying physics 
choices, and also discuss the difficulties in comparing photon pdfs in 
leading order (LO) and next-to-leading order (NLO) pQCD.

These remarks only apply to real photons. For virtual photons, there is 
very little experimental information from two photon collisions, because
of the difficulties of doing double-tag measurements. However, there 
have been some theoretical attempts, which we discuss in sect.\ 3. 

Hard photoproduction at HERA offers further possibilities of exploring 
the structure of the photon, including the gluon content. In hard 
photoproduction processes, there is a contribution from the {\bf 
resolved} processes, where the photon is resolved into its partons which
then take part in the hard partonic subprocess \cite{Kramer,JKS}: this 
contribution is sensitive to the pdfs of the photon as well as the 
proton. There is also a (calculable) background to this from the {\bf 
direct} processes, in which the photon takes part directly in the hard 
subprocess: this depends on the proton pdfs but not on those of the 
photon. 

With untagged electrons at HERA the photons are mainly real and have a 
known spectrum of energies given by the equivalent photon approximation;
hence the pdfs of the real photon are measured. With tagged electrons 
the photons have known energy and virtuality and so it will become 
possible to study the parton content of virtual photons for the first 
time. This is why the structure of real and virtual photons is an 
important physics issue for HERA. 

{\bf 2. The parton distributions of the real photon}

In this section we briefly review the different sets of pdfs for the 
real photon currently available: for more detail on this subject the 
reader is referred to ref.~\cite{Vogt}. For a recent general review see 
ref.~\cite{DG95}. The special role of the photon in QCD is due to the 
fact that, at asymptotically large Q$^{2}$, the quark and gluon 
distribution functions are calculable at large $x$, i.e.
\begin{equation}
q^{\gamma}_{i} (x, Q^{2})/\alpha \ \simeq \ \frac{a_{i}(x)}{\alpha_{s}
(Q^{2})} \ + \ b_{i}(x) \:\:\: ,
\end{equation}
with a similar expression for $g^{\gamma} (x,Q^{2})$. The first term 
is the LO result of Witten \cite{Witten} and the second term its NLO
correction in pQCD \cite{Bard}. The functions $a_{i}(x)$ and $b_{i}(x)$ 
are {\bf calculable}, but singular at $x$ = 0.

This point-like part contribution is dominant at large $Q^2$, where 
the incalculable hadronic piece is small. However to avoid the 
unphysical small-$x$ singularities in eq.~(1) one must retain the 
hadronic part by including a boundary condition at a reference scale 
$Q^{2} = Q^{2}_{0}$ \cite {GGR}. If we do that in $n$-moment space we 
have (confining ourselves to LO for simplicity) 
\begin{equation}
q^{\gamma}_{i} (n, Q^{2}) = \frac{\alpha\, a_{i} (n)}{\alpha_{s}(Q^{2})}
\left[ 1 - \left( \frac{\alpha_{s} (Q^{2})}{\alpha_{s} (Q^{2}_{0})} 
\right) ^{1 + d(n)} \right] + q^{\gamma}_{i} (n, Q^{2}_{0}) \left( 
\frac{\alpha_{s} (Q^{2})}{\alpha_{s} (Q^{2}_{0})} \right) ^{d(n)}
\end{equation}
where the second term in the square brackets regularizes the singularity
at $x$ = 0. The last term is hadronic in the sense that for the case 
of the pdfs of a hadron, it would be the {\bf only} contribution.

Eq.~(2), although a good approximation at large $x$, is strictly true 
only for non-singlet combinations of quark densities. The singlet quark 
$\Sigma^{\gamma} = \sum_{i} (q_{i} + {\overline q}_{i}) $ and gluon 
distributions obey similar equations with the important difference 
that in these sectors the Altarelli-Parisi (AP) equations are {\bf 
coupled}, i.e.\ to determine $\Sigma^{\gamma} (x, Q^{2})$ (or 
$g^{\gamma} (x, Q^{2}))$ we need both $\Sigma^{\gamma} (x, Q^{2}_{0})$ 
{\bf and} $g^{\gamma} (x, Q^{2}_{0})$.  The bottom line is that, as 
for the case of hadrons, we need input distributions at a reference 
scale $Q^{2} = Q^{2}_{0}$ .
 
We note here that the anomalous [$\alpha _{s} (Q^{2})]^{-1} $ behaviour 
of the pdfs of the photon arises because of the direct $\gamma 
\rightarrow q {\overline q} $ coupling which gives inhomogeneous terms 
in the AP evolution equations \cite{Evol}. This behaviour of the quark 
distributions is confirmed by data on $F^{\gamma}_{2} (x, Q^{2})$, over 
a wide range of $Q^{2}$, see e.g.\ ref.~\cite{f2data}. 

In order to obtain photon pdfs at all $Q^2$, one has to choose a 
reference scale $Q_0^2$ and fix the input pdfs there, using some ansatz 
(usually vector meson dominance, VMD) and employing $F^{\gamma}_{2}$ 
data to fix some free parameters. The (anti)quark distributions 
$q^{\gamma}_{i} (x, Q_0^{2})$ are reasonably well determined by the 
data, since in LO 
\begin{equation}
F^{\gamma}_{2} (x, Q^{2}) \ = \ \sum_{i} e^{2}_{i}\, xq^{\gamma}_{i} 
(x,Q^{2}) \:\:\: .
\end{equation}
On the other hand, fixing the gluon distribution is a problem because 
of the lack of a momentum sum rule \cite{Vogt,JKSLund}: one is 
completely dependent on the ansatz. Note however that a substitute for 
the usual hadronic parton momentum sum rule has been proposed recently 
\cite{Frankf}. For the evolved pdfs the fact that the coupling of the 
AP equations is weak works two ways: (a) the output $F^{\gamma}_{2} 
(x,Q^2)$ is insensitive to the input gluon (except at small $x$) but 
(b), consequently, a comparison with present $F_2^{\gamma}$ data does 
not provide any restriction on $g^{\gamma}(x,Q_0^2)$.

VMD provides a connection between the photon and $\rho $ meson pdfs, 
and since the latter satisfy a momentum sum rule, we have a constraint 
on the VMD part of the gluon pdf of the photon. This is particularly 
useful if we use SU(6) to relate the pion and $\rho $ pdfs as there are 
experimental constraints on the pion pdfs from Drell-Yan lepton pair 
and direct-photon production data \cite{pipdfs}. However, another 
problem arises here. For the traditional input scales, $Q_0^2\ge 1$
GeV$^2$, a pure VMD input is known to be insufficient to fit the data 
at higher $Q^2$ \cite{GGR,Jan}. Two approaches have been adopted to 
circumvent this: the first is to maintain the VMD idea and start the AP 
evolution at a very low scale $Q_0 < 1$ GeV \cite{GRV,AFG1,AFG2,SaS}. 
The second is to keep $Q_0 \ge 1$ GeV and fit the quark densities to 
$F_2^\gamma $ data, e.g.\ supplementing the VMD values with a point-like
component, seemingly naturally provided by the Born-Box diagram. 
Unfortunately there is no corresponding natural choice for the gluon 
density and a guess must be made here. This method was adopted in
refs.\ \cite{SaS,Dgrassie,LAC,GS,WHIT}. The result of all this is that 
the different distributions agree reasonably well as regards the quark 
distributions in the region $0.05 \lessim x \lessim 0.8$, which must 
reproduce $F_2^\gamma $ data, but not as regards the gluon densities. 
This can be seen in figs.\ 1 and 2, where we have plotted a 
representative set of quark and gluon distributions in LO and NLO. 
\begin{figure}[thb]
\begin{center}
\vspace*{-0.5cm}
\epsfig{file=ugamma.eps,height=9.8cm}
\vspace*{-1.2cm}
\end{center}
\caption[]
{Photonic $u$-quark parametrizations at LO \protect\cite{GRV,SaS,LAC,GS}
 and NLO \protect\cite{GRV,AFG2,GS}. The NLO results are presented in 
 the $\overline{\mbox{MS}}$ scheme. For a discussion of the different 
 assumptions and $F_{2}^{\,\gamma}$ data sets employed see \protect\cite
 {Vogt}.}
\end{figure}
\begin{figure}[thb]
\begin{center}
\vspace*{-0.5cm}
\epsfig{file=ggamma.eps,height=9.8cm}
\vspace*{-1.2cm}
\end{center}
\caption[]
{Parametrizations of the photon's gluon distribution at LO \protect\cite
 {GRV,SaS,LAC,GS} and NLO \protect\cite{GRV,AFG2,GS}. Note that the 
 similarity of the NLO results is due to common VMD prejudices and not 
 enforced by data.}
\end{figure}

The reader will note that there seems to be very little resemblance 
between the quark distributions in LO and NLO. This is because of a 
subtlety peculiar to the photon in the $\overline{\mbox{MS}}$ scheme,
and has been discussed in detail in ref.\ \cite{Vogt}. It arises in the 
lowest order QCD process, the Born-Box diagram, where the term leading 
in $\ln Q^2 $ gives the inhomogeneous term in the AP equations and the 
non-leading term $C_{\gamma}$ is negative and divergent as $x\to 1$. 
In the usual $\overline{\mbox{MS}}$ scheme this $C_{\gamma}$ is not 
absorbed into the quark densities. However, in NLO, it reappears as a 
Wilson coefficient for a subleading `direct' contribution to $F_2
^\gamma$. Thus if we are to require approximately the same $F_2^\gamma$
in LO and NLO, then the NLO pdfs must be substantially modified 
accordingly and this is what we are seeing in fig.\ 1. An alternative 
approach is to work in the $\mbox{DIS}_\gamma$ scheme \cite{GRV2}, where
$C_\gamma$ is absorbed into the definition of the quark density and does
not appear in the NLO expression for $F_2^\gamma$. Hence in this scheme 
perturbatively stable, physically motivated inputs for the photon pdfs, 
such as VMD, can be used in NLO as well as in LO. 

We conclude this section with a few comments on what has been learnt 
from experimental data since most of these pdfs were proposed. We start
with two-photon data. There have been new $F_2^\gamma$ data from TOPAZ 
\cite{TOPAZ1} and AMY \cite{AMY1} at TRISTAN and from OPAL \cite{OPAL} 
and DELPHI \cite{DELPHI} at LEP, which are shown in fig.\ 3. As can 
be seen the data are of limited statistics. These results seem to 
indicate some offset at $x$ around 0.2 with respect to the average of 
earlier data from lower energy machines, as can be seen by comparing 
to the LAC \cite{LAC} and GRV \cite{GRV} parametrizations which were 
fitted to all $F_2^\gamma$ data available in 1991. Moreover, at small 
$x$ the recent TOPAZ \cite{TOPAZ1} results are at variance with the 
LEP data \cite{OPAL,DELPHI}. If anything, the recent measurements 
confuse the situation slightly as regards the quark distributions. 
In addition, there have been measurements of the one- and two-jet 
inclusive jet cross sections at TRISTAN \cite{TOPAZ2,AMY2} which show 
some sensitivity to the gluon distribution. One can conclude from these 
jet data that there is now evidence from $\gamma \gamma $ collisions 
that the gluon density is non-zero \cite{JKSLund,TOPAZ2,AMY2}. They also
rule out the LAC3 distribution with its large gluon component at large 
$x$, which considerably overestimates the cross section.
\begin{figure}[thb]
\begin{center}
\vspace*{-0.8cm}
\epsfig{file=f2gam.eps,height=12cm}
\vspace*{-1.2cm}
\end{center}
\caption[]
{Recent data on $F_{2}^{\gamma}(x,Q^2)$ from TRISTAN \protect\cite
 {TOPAZ1,AMY1} and LEP \protect\cite{OPAL,DELPHI} compared to the LO 
 fits to all previous data of LAC \protect\cite{LAC} and GRV 
 \protect\cite{GRV}.}
\end{figure}

Turning to jet photoproduction at HERA, already the first measurements 
\cite{H1old} enabled LAC3 to be ruled out: the contribution from the 
quarks virtually saturates the observed cross section, leaving no scope 
for a large gluon density except at small $x$ \cite{JKSLund}. Since 
then more accurate data have appeared. H1 has extracted a LO gluon 
distribution from the data \cite{H1new}: it disfavours the more extreme 
scenarios for gluons at small-$x$ such as LAC1 and LAC2. These latter 
data have been compared with NLO calculations, as discussed at this 
workshop \cite{Kramer}. It appears that in the negative rapidity region,
where originally the direct component was expected to dominate, there 
is some sensitivity to the large-$x$ photon pdfs. A large $x$ quark 
structure more in accord with the GS distributions than those of GRV or 
AFG seems to be favoured, although some theoretical questions have to 
be answered before definite conclusions can be drawn \cite{Kramer}. 
In the positive rapidity region, a comparison of the NLO calculations 
with the jet cross sections is complicated by the possibility of 
multiple hard parton interactions, discussed in a separate contribution
to these proceedings \cite{Multip}.   

{\bf 3. The parton distributions of the virtual photon}

As with the pdfs of the real photon, the pdfs of a virtual photon 
have to be based on some ansatz. There is therefore no unique answer.
The non-perturbative hadronic (VMD) contribution to the photon
structure is expected to go away with increasing $P^2$, allowing for a
purely perturbative prediction for $F_2^{\,\gamma}(x,Q^2,P^2)$ at
sufficiently high $P^2$ \cite{uematsu}. Here we use $P^2$ to denote the 
virtuality of the photon; $Q^2$ is reserved for the scale of the hard 
$\gamma^{(*)}$ interaction. The fall-off of the non-perturbative part 
with increasing $P^2$ is theoretically uncertain; hence experimental 
clarification is required to pin down models. We will summarize here a 
few recent studies that together illustrate the spread in current 
approaches.

Drees and Godbole \cite{DG} seek a simple interpolating multiplicative
factor, such that parton distributions reduce to the real pdfs for 
$P^2 \to 0$ and die like $\ln(Q^2/P^2)$ for $P^2 \to Q^2$: at $P^2 = 
Q^2$ it is natural to attribute the whole cross section to direct 
processes in order to avoid double counting. Several different forms 
are studied; one of the main alternatives is to use a scaling factor:
\begin{equation}
r = 1 - \frac{\ln(1+P^2/P_c^2)}{\ln(1+Q^2/P_c^2)} ~,
\end{equation}  
where $P_c$ is some typical hadronic scale such as $P_c^2 \approx 
0.5$~GeV$^2$. The factor $r$ is applied to all quark pdfs. The gluon is 
expected to be further suppressed, however, since the gluon pdf is 
generated by the quark ones \cite{Borzu}. For instance, if the scale 
$k^2$ of $\gamma \to \mathrm{q}\overline{\mathrm{q}}$ branchings is 
distributed in the range $P^2 \lessim k^2 \lessim Q^2$, the scale 
$k'^2$ of the q~$\to$~qg branching is in the reduced range $k^2 \lessim 
k'^2 \lessim Q^2$. A gluon suppression factor $r^2$ gives the expected 
limiting behaviour. 
The above ansatz does not change the $x$ shape of distributions;
for that more complicated forms are proposed. Anyway, the thrust 
of the study is to estimate how much the photon pdfs in the untagged 
case, i.e.\ $P^2$-averaged pdfs, differ from those of the real photon. 
The forms studied give a suppression of the order of 10\% and 15\%
for the quark and gluon distributions, respectively.  

The study of Gl\"uck, Reya and Stratmann \cite{GRS} is based on the
observation that the pdfs $f_i^{\gamma}(x, Q^2, P^2)$ obey 
evolution equations in $Q^2$ similar to those of a real photon. 
The question is therefore reduced to one of finding suitable boundary 
conditions at $Q^2 = P^2$. The ansatz used is
\begin{equation}
f_i^{\gamma}(x, Q^2= \tilde{P}^2, P^2) = 
\eta(P^2) f_i^{\gamma,\mathrm{nonpert}}(x, \tilde{P}^2) +
\left[ 1-\eta(P^2) \right] f_i^{\gamma,\mathrm{pert}}(x, \tilde{P}^2)
~.
\end{equation}   
Here $\tilde{P}^2 = \max(P^2,\mu^2)$, with $\mu \approx 0.5$~GeV
the input scale for the evolution of the real photon \cite{GRV};
and $\eta(P^2) = (1 + P^2/m_{\rho}^2)^{-2}$ is the standard
dipole dampening factor of the $\rho$ meson. The non-perturbative
input distribution is taken to be proportional to the pfd's of the 
pion, $f_i^{\gamma,\mathrm{nonpert}}(x, \tilde{P}^2) =
\kappa (4\pi\alpha/f_{\rho}^2) f_i^{\pi}(x, \tilde{P}^2)$ \cite{GRV}.
The $f_i^{\gamma,\mathrm{pert}}(x, \tilde{P}^2)$ is perturbatively
calculable; in leading order it vanishes.
Based on the above ansatz, the evolution equations give the answer for
all $Q^2 > P^2$. Closed results can be obtained in moment space, and 
then a simple numerical Mellin inversion gives actual numbers. 
A practical limitation is that there exists up to now no simple 
parametrization, unlike the case of a real photon. 

Schuler and Sj\"ostrand \cite{SaS} start from an ansatz for the pdfs 
of a real photon decomposed into VMD and anomalous components:
\begin{equation}
f_i^{\gamma}(x, Q^2) = \sum_V \frac{4\pi\alpha}{f_V^2} 
f_i^{\gamma,V}(x, Q^2, Q_0^2) + 
\int_{Q_0^2}^{Q^2} \frac{{\mathrm{d}}k^2}{k^2}
\frac{\alpha}{2\pi} \sum_{\mathrm{q}} 2e_{\mathrm{q}}
f_i^{\gamma,\mathrm{q}\overline{\mathrm{q}}}(x, Q^2, k^2) ~.
\label{SaSansatz}
\end{equation}
Here the sum runs over the lowest-lying vector mesons, $\rho^0$,
$\omega$, $\phi$ and J/$\psi$, while the integral covers the range of
perturbative branchings $\gamma\to\mathrm{q}\overline{\mathrm{q}}$ 
at scales $Q_0 < k < Q$, with $Q_0 \approx 0.6$~GeV (for SaS 1,
alternatively 2~GeV for SaS 2) setting the separation between the 
two components and also the starting value of the evolution. The VMD 
and anomalous ``state'' distributions $f_i^{\gamma,V}$ and 
$f_i^{\gamma,\mathrm{q}\overline{\mathrm{q}}}$, respectively, are 
normalized to unit momentum sum.
The VMD distributions and the integral of anomalous distributions
are parametrized separately and added to give the full result. In going 
to a virtual photon, the main change is to introduce a dipole dampening
factor for each component, i.e. $(1 + P^2/m_V^2)^{-2}$ for the VMD
states and $(1 + P^2/k^2)^{-2}$ for the anomalous ones. Additionally
the lower input scale for the VMD states is shifted from $Q_0^2$
to $P_0^2 \approx \max(P^2,Q_0^2)$ \cite{Borzu}. 

In order to obtain a tractable answer, one possible approximation 
for the anomalous component is
\begin{equation}
\int_{Q_0^2}^{Q^2} \frac{1}{(1+P^2/k^2)^2} \frac{{\mathrm{d}}k^2}{k^2}
\Big[ \cdots \Big] \approx \int_{P_0^2}^{Q^2} 
\frac{{\mathrm{d}}k^2}{k^2} \Big[ \cdots \Big] ~,
\label{SaSapprox}
\end{equation}
with $P_0$ as above. Although the VMD and anomalous components
still depend on two scales, $P_0^2$ and $Q^2$, all the nontrivial
dependence comes from the logarithmic integration of the strong 
coupling constant between the two scales, so the standard pdfs 
of the real photon can be extended easily to virtual photons, 
i.e. parametrizations of $f_i^{\gamma}(x, Q^2, P^2)$ are 
readily available. The resulting $u$-quark and gluon densities are
dispayed for two photon virtualities $P^2$ in fig.~4 together with
the corresponding LO distributions of ref.\ \cite{SaS}.
Recently, alternatives to eq.~(\ref{SaSapprox}) have been studied 
\cite{SaSprep}, where the momentum sum and average evolution range 
of the dipole-dampened version of eq.~(\ref{SaSansatz}) is preserved. 
The difference between these procedures can also be viewed as one 
estimate of the uncertainty.  
\begin{figure}[thb]
\begin{center}
\vspace*{-0.8cm}
\epsfig{file=ugvirt.eps,height=9.8cm}
\vspace*{-1.2cm}
\end{center}
\caption[]
{The $u$-quark and the gluon distributions of the virtual photon in LO 
 as suggested in refs.\ \protect\cite{SaS,GRS} at two selected values
 of the photon virtuality $P^2$.}
\end{figure}
 
We now turn to the experimental possibilities at HERA. With their 
forward electron tagging capabilities, H1 and ZEUS can tag almost real 
photons, $P^2 \lessim 0.01$~GeV$^2$, and virtual photons down to $P^2 
\gtrsim 0.1$~GeV$^2$. This is amply demonstrated by the ZEUS results 
presented at this meeting \cite{ZEUS}, shown in fig.7 of ref.\ \cite
{Maxfield}, where the observed $x_{\gamma}$ distribution has been 
constructed for events with two jets above 4 GeV, i.e.\ roughly 
speaking for events with $Q^2 \gtrsim 16$~GeV$^2$.
As $P^2$ is increased, this distribution is gradually suppressed at
small $x_{\gamma}$, where the resolved contribution should dominate. 
For instance, if we cut between resolved and direct events at 
$x_{\gamma} = 0.75$, then the ratio of resolved events (with $x_{\gamma}
 < 0.75$) to direct events (with $x_{\gamma} > 0.75$) drops by about a 
factor of 2 between $P^2 \approx 0$ and $P^2 \approx 0.5$~GeV$^2$ \cite
{Maxfield}, in rough agreement with the theoretical arguments of this 
section. 
 
{\bf 4. Conclusions}
        
In this note we have briefly reviewed the current phenomenological 
status of the pdfs of real and virtual photons. As we mentioned earlier,
in the case of the real photon the pdfs have been constrained to fit the
$F_2^\gamma $ data. However, given the limitations catalogued in sect.\
2, it is difficult to regard a comparison of HERA jet photoproduction 
data with NLO QCD calculations based on the existing pdfs as a 
definitive test of anything. We feel that the jet data should be 
regarded as giving an independent determination of the pdfs of the 
photon, which at the moment is and in the near future will remain 
superior to those from two-photon physics. This will be true, certainly 
for larger $x$, until high statistics data become available from LEP2. 
We are already seeing the first signs of this \cite{Kramer} in the 
comparison of the jet data with NLO calculations. It should also be 
borne in mind, that jet studies (in both $\gamma \gamma $ and $\gamma$p)
are among the few areas sensitive to the gluon content of the photon. 
Thus we feel that photoproduction of jets at HERA has much to offer the 
field of photon structure studies.

As regards virtual photons, HERA offers a unique opportunity to study 
pdfs. The electron tagging capabilities of H1 and ZEUS offers virtual 
photons with large event rates. As this area has been regarded to be of 
theoretical interest for nearly 20 years, but with essentially no 
experimental input, this is an opportunity not to be missed. In the 
future, LEP2 may provide complementary information on virtual photons 
via double-tag events, but at a lower energy and presumably with lower 
event rates \cite{LEP2report}.
\vspace{-0.5cm}

\begin{thebibliographyss}{99}

\bibitem{Witten}
E. Witten, Nucl.\ Phys.\ {\bf B120} (1977) 189.

\bibitem{Bard}
W.A. Bardeen and A.J. Buras, Phys.\ Rev.\ {\bf D20} (1979) 166; 
{\bf D21} (1980) 2041(E). 

\bibitem{Kramer}
G.Kramer, these proceedings.

\bibitem{JKS}
J.K. Storrow, J.\ Phys.\ {\bf G19} (1993) 1641.

\bibitem{Vogt}
A. Vogt, Proceedings of the Workshop on Two-Photon Physics at LEP and 
HERA, Lund, May 1994, eds.\ G.Jarlskog and L.J\"onsson (Lund Univ., 
1994), p.\ 141.

\bibitem{DG95}
M. Drees and R.M. Godbole, Univ.\ Wisconsin preprint MAD-PH-898 (1995). 

\bibitem{GGR} 
M. Gl\"uck, K. Grassie and E. Reya, Phys.\ Rev.\ {\bf D30} (1984) 1447.

\bibitem{Evol} R.J. De Witt et al., Phys.\ Rev.\ {\bf D19} (1979) 2046;
{\bf D20} (1979) 1751(E).

\bibitem{f2data} 
D.J. Miller, Proceedings of the Workshop on Two-Photon Physics at LEP 
and HERA, Lund, May 1994, eds.\ G.Jarlskog and L.J\"onsson (Lund Univ., 
1994), p.\ 4.

\bibitem{JKSLund} 
J.K. Storrow, Proceedings of the Workshop on Two-Photon Physics at LEP
and HERA, Lund, May 1994, eds.\ G.Jarlskog and L.J\"onsson (Lund Univ., 
1994), p.\ 149.

\bibitem{Frankf} 
L.L. Frankfurt and E.G. Gurvich, these proceedings.

\bibitem{pipdfs} 
P.\ Aurenche et al, Z.\ Phys.\ {\bf C56} (1992) 589; \\
M. Gl\"uck, E. Reya and A.\ Vogt, Z.\ Phys.\ {\bf C53} (1992) 651; \\
P.J. Sutton et al, Phys.\ Rev.\ {\bf D45} (1992) 2349. 

\bibitem{Jan} 
J.H. Da Luz Vieira and J.K. Storrow, Z.\ Phys.\ {\bf C51} (1991) 241.
 
\bibitem{GRV}
M. Gl\"uck, E. Reya and A. Vogt, Phys.\ Rev.\ {\bf D46} (1992) 1973.

\bibitem{AFG1} 
P.Aurenche et al, Z.\ Phys.\ {\bf C56} (1992) 589.

\bibitem{AFG2}
P. Aurenche, M. Fontannaz and J.P. Guillet, Z.\ Phys.\ {\bf C64} (1994) 
621. 

\bibitem{SaS}
G.A. Schuler and T. Sj\"ostrand, Z.\ Phys.\ {\bf C68} (1995) 607.

\bibitem{Dgrassie}
M. Drees and K. Grassie, Z.\ Phys.\ {\bf C28} (1994) 451.

\bibitem{LAC} 
H. Abramowicz, K. Charchula and A. Levy, Phys.\ Lett.\ {\bf B269} (1991)
458.

\bibitem{GS} 
L.E. Gordon and J.K. Storrow, Z.\ Phys.\ {\bf C56} (1992) 307.

\bibitem{WHIT} 
K. Hagiwara et al, Phys.\ Rev.\ {\bf D51} (1995) 3197.

\bibitem{GRV2}
M. Gl\"uck, E. Reya and A. Vogt, Phys.\ Rev.\ {\bf D45} (1992) 3986.  

\bibitem{TOPAZ1} 
K.Muramatsu et al (TOPAZ), Phys.\ Lett.\ {\bf B332} (1994) 477.
 
\bibitem{AMY1} 
S.K.Sahu et al (AMY), Phys.\ Lett.\ {\bf B346} (1995) 208.

\bibitem{OPAL}
R.Akers et al (OPAL), Z.\ Phys.\ {\bf C61} (1994) 199.
 
\bibitem{DELPHI}
P.Abreu et al (DELPHI), CERN preprint PPE-95-87.
 
\bibitem{TOPAZ2}
H.Hayashii et al (TOPAZ), Phys.\ Lett.\ {\bf B314} (1993) 149. 

\bibitem{AMY2}
B.J.Kim et al (AMY), Phys.\ Lett.\ {\bf B325} (1994) 248. 

\bibitem{H1old}
I. Abt et al (H1) Phys.\ Lett.\ {\bf B314} (1993) 436; J. Dainton, 
16th Int.\ Symp.\ on Lepton-Photon Interactions (1993). 

\bibitem{H1new}
T.Ahmed et al (H1), Nucl.\ Phys.\ {\bf B445} (1995) 195. 

\bibitem{Multip}
J.M. Butterworth et al, these proceedings.

\bibitem{uematsu} T. Uematsu and T.F. Walsh, Phys.\ Lett.\ {\bf B101}
(1981) 263, Nucl.\ Phys.\ {\bf B199} (1982) 93; \\
G. Rossi, Univ.\ California report UCSD-10P-227 (1983), 
Phys.\ Rev.\ {\bf D29} (1984) 852.

\bibitem{DG} 
M. Drees and R.M. Godbole, Phys.\ Rev.\ {\bf D50} (1994) 3124.

\bibitem{Borzu}
F.M. Borzumati and G.A. Schuler, Z.\ Phys.\ {\bf C58} (1993) 139. 

\bibitem{GRS}
M. Gl\"uck, E. Reya and M. Stratmann, Phys.\ Rev.\ {\bf D51} (1995)
3220.

\bibitem{SaSprep}
G.A. Schuler and T. Sj\"ostrand, CERN--TH/96--04 and LU TP 96--2.

\bibitem{ZEUS}
M. Utley (ZEUS), private communication.

\bibitem{Maxfield}
S. Maxfield, these proceedings.

\bibitem{LEP2report}
P. Aurenche, G.A. Schuler et al., ``$\gamma\gamma$ Physics'',
to appear in the proceedings of the 1995 LEP~2 Physics workshop.

\end{thebibliographyss}

\end{document}